\documentclass[10pt,letterpaper,twocolumn]{article} 
\usepackage{ol2}

\usepackage{amsmath}

\begin{document}

\twocolumn[ 

\title{Polychromatic Optical Bloch Oscillations}


\author{Stefano Longhi}
\address{Dipartimento di Fisica and Istituto di Fotonica e Nanotecnologie del CNR,
Politecnico di Milano, Piazza L. da Vinci 32, I-20133 Milano, Italy}
\begin{abstract}
Bloch oscillations (BOs) of polychromatic beams in circularly-curved optical waveguide arrays
are smeared out owing to the dependence of the BO spatial period on wavelength. Here it is shown that
restoring of the self-imaging property of the array and approximate BOs over
  relatively broad spectral ranges can be achieved by insertion of suitable lumped phase slips uniformly applied across the array.

\end{abstract}



] 

\noindent

Photonic lattices have provided in recent years a laboratory tool to
visualize optical analogues of coherent phenomena generally
encountered in solid-state physics, such as Bloch oscillations (BOs)
\cite{Morandotti99,Pertsch99,Christodoulides03,Sapienza03} and
dynamic localization (DL) \cite{Longhi06,Iyer07} in lattices driven
by external dc or ac fields. In waveguide arrays, a common
engineering approach to realize optical BOs and DL  is to tailor the
local curvature of the array axis \cite{Lenz99,Longhi05}. DL and BOs
have been observed in bent waveguide arrays with either
periodically-varying or constant curvature to mimic ac or dc fields
\cite{Longhi06,Iyer07,Chiodo06,Dreisow08}. In optics, BOs and DL
enable light diffraction management, with the possibility of
periodic image reconstruction \cite{Longhi05,Eisenberg00}. Such a
self-imaging effect, however, is generally limited to relatively
narrow spectral beams. In case of DL, destruction of self-imaging
for broadband beams stems from the resonant nature of the DL
condition \cite{Dunlap86}. Here, periodic self-imaging is exactly
fulfilled solely for a target wavelength, whereas other spectral
components experience residual diffraction. In \cite{Garanovich06},
Garanovich and coworkers suggested special profiles of waveguide
axis curvature that mix first and second resonances of DL, leading
to approximate DL  over an extremely broad spectral region. The
experimental demonstration of broadband DL has been very recently
reported by Szameit and collaborators \cite{Szameit09}. In case of
BOs, destruction of exact self-imaging with polychromatic beams has
a different reason. Here periodic beam reconstruction is exactly
attained for each spectral beam component, however the self-imaging
period, given by the Bloch period $s_B$, turns out to be
proportional to the wavelength $\lambda$: therefore, for a beam
spectrally broadened by less than $\sim 10 \%$ around its carrier,
field reconstruction
is fully lost just after few BO cycles.\\
In this Letter it is shown that polychromatic optical BOs can be
approximately achieved over a broad spectral range by suitable
insertion of lumped phase slips in the array, which combat the
dispersion of BO period with wavelength. Let us consider light
propagation in low-contrast-index one-dimensional waveguide arrays
with a weakly curved axis in the $(x,z)$ plane. In the curvilinear
coordinate system $(s,\eta)$ of Fig.1(a), where $s$ is the arc
length of the curved waveguide axis, the propagation for the
spectral field component $\psi(\eta,s; \lambda)$ at wavelength
$\lambda$ is described by the Schr\"{o}dinger-like equation
\begin{equation}
i \hbar \frac{\partial \psi}{\partial s}=-\frac{\hbar ^2}{2 n_s}
\frac{\partial^2 \psi}{\partial \eta^2}+[n_s-n(\eta)] \psi + n_s
\kappa(s) \eta \psi
\end{equation}
\begin{figure}[htb]
\centerline{\includegraphics[width=8.2cm]{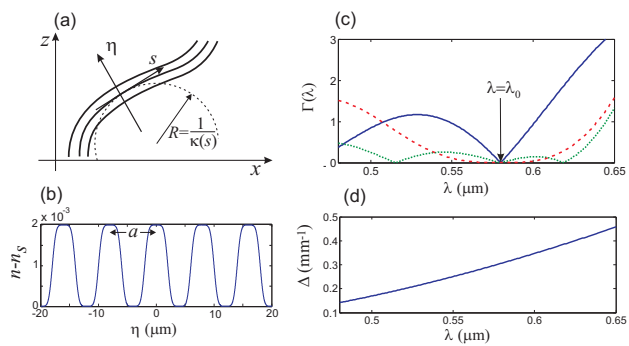}} \caption{
(Color online) (a) Schematic of a curved waveguide array. (b)
Refractive index profile $n(\eta)-n_s$ of the array used in
numerical simulations ($a=8 \; \mu$m, $n_s=1.45$). (c) Behavior of
$\Gamma(\lambda)$, entering  in Eq.(3), versus wavelength for (i)
the 8-cm-long circularly-curved array without lumped phase slips
(solid curve); (ii) $\alpha_1=\alpha_3=\lambda_0/(2 a n_s)$ and
$\alpha_2=0$ (dashed curve); (iii) for
$\alpha_1=\alpha_2=\alpha_3=\lambda_0/(3 a n_s)$ (dotted curve). (d)
Behavior of waveguide coupling rate $\Delta$ versus wavelength.}
\end{figure}
 where $\hbar= \lambda / (2 \pi)$ is the reduced wavelength, $n_s$ is
the bulk refractive index, $n(\eta)$ is the periodic index profile
of the array with lattice period $a$, and $\kappa(s)$ is the local
curvature of waveguide axis. In the single-band and nearest-neighbor
tight-binding approximations, Eq.(1) reduces to standard coupled
mode equations for amplitudes $c_n(s; \lambda)$ of modes trapped in
the $n$-th waveguide of the array  \cite{Longhi05,Garanovich06}
\begin{equation}
i \dot{c}_n=-\Delta (\lambda) (c_{n+1}+c_{n-1})+\frac{2 \pi n_s
\kappa(s)an}{\lambda}c_n
\end{equation}
where $\Delta(\lambda)$ is the coupling rate, at wavelength
$\lambda$, between adjacent waveguides, and the dot denotes the
derivative with respect to the arc length. Here we consider a
waveguide array with constant curvature $\kappa=1/R$, into which a
sequence of lumped phase gradients are superimposed at the arc
lengths $s_1$, $s_2$,...., i.e. we assume in Eq.(2)
$\kappa(s)=1/R+\alpha_1 \delta(s-s_1)+\alpha_2 \delta(s-s_2)+...$,
where the parameters $\alpha_1$, $\alpha_2$,... define the phase
slips $\varphi_1$, $\varphi_2$,... introduced uniformly across the
array at the arc lengths $s_1$, $s_2$,.... The relation between the
phase slip $\varphi_l$ and the parameter $\alpha_l$ is simply
obtained after integration of Eq.(2) in the infinitesimal interval
$(s=s_l^-$,$s=s_l^+)$, yielding $\varphi_l=2 \pi n_s a \alpha_l/
\lambda$. In practice, a lumped phase gradient can be introduced (i)
by a sudden tilt of waveguide axis \cite{Eisenberg00} by a small
angle $\theta$; in this case $ \alpha_l=\theta$ and $\varphi_l=\pi
\theta/ \theta_B$, where $\theta_B=\lambda/(2 a n_s)$ is the Bragg
angle; (ii) by local waveguide segmentation, channel narrowing or
index change modulation \cite{Szameit08}. To investigate the
self-imaging property of the arrayed structure, it is enough to
consider the impulse response of the array corresponding to
excitation of the waveguide $n=0$. The solution to Eq.(2) with the
initial condition $c_n(0)=\delta_{n,0}$ reads \cite{Dunlap86}
\begin{equation}
|c_n(s)|^2 =J_n^2\left(2 \Gamma(\lambda) \right),
\end{equation}
where $J_n$ is the Bessel function of order $n$,
$\Gamma(\lambda)=\Delta(\lambda) |q(s,\lambda)|$ and
\begin{equation}
q(s,\lambda)=\int_0^s d \xi \exp[-i \gamma(\xi) / \lambda], \;
\gamma(s)=2 \pi n_s a  \int_0^s d \xi \kappa(\xi).
\end{equation}
The condition for self-imaging after a propagation length $s$,  i.e.
$|c_n(s)|^2= \delta_{n,0}$, is thus $q(s,\lambda)=0$. For the BO
problem in absence of phase slips, $\kappa(s)=1/R$  and
$|q(s,\lambda)|=s |\sin( \pi s/s_B)|/(\pi s/s_B)$, where
\begin{equation}
s_B(\lambda)= \lambda R/(n_s a)
\end{equation}
is the BO period for the spectral field component of wavelength
$\lambda$. For a monochromatic beam with carrier wavelength
$\lambda_0$, periodic self-imaging is thus attained at propagation
lengths $s$ which are integer multiples of the BO cycle
$s_B(\lambda_0)$. Unfortunately, for spectrally-broad beams
different wavelength components $\lambda$ show a shifted BO cycle
$s_B(\lambda)$, making self-imaging and polychromatic BOs to rapidly
washing out for a relatively broad spectrum of the incoming beam. As
an example, Fig.2(a) shows the breakdown of self-imaging and
smearing out of a breathing BO mode for polychromatic light in a
$L=8$-cm-long circularly-curved array as obtained by numerical
simulations of Eq.(1) for an index profile of the array shown in
Fig.1(b) and for a radius $R=40$ cm. The intensity distribution of
Fig.2(a) is  obtained by the incoherent superposition
 of five monochromatic fields of equal power (at $\lambda=$540, 560, 580, 600, and 620
 nm), which  excite simultaneously the waveguide $n=0$.
For comparison, Fig.2(b) shows the evolution of beam intensity for
the carrier spectral component solely ($\lambda=\lambda_0=580$ nm),
which undergoes periodic self imaging with spatial period
$s_B(\lambda_0)=\lambda_0 R/(n_s a)=2$ cm. Note that the full
propagation length $L$ cm comprises exactly $N=4$ BO cycles at the
carrier wavelength. The washing out of self imaging at the output
plane for a spectrally broad beam is related to the rapid deviation
from zero of $\Gamma(\lambda)$ as $\lambda$ varies by a a few
percents around the reference wavelength $\lambda_0$, as shown by
the solid curve in Fig.1(c). The allowed spectral band $\Delta
\lambda$ for (approximate) self-imaging can be roughly estimated
from the condition $2|\Gamma(\lambda)| < 1$ [see Eq.(3)], and turns
out to be $\Delta \lambda \sim 20$ nm for the solid curve of
Fig.1(c). The introduction of appropriate phase slips across the
array may be exploited to flatten the function $\Gamma(\lambda)$ at
around $\lambda= \lambda_0$. To this aim, let us consider $(N-1)$
lumped phase gradients at $s_1=s_B(\lambda_0)$,
$s_2=2s_B(\lambda_0)$,...,$s_{N-1}=(N-1)s_B(\lambda_0)$, i.e. let us
assume $\kappa(s)=1/R+\sum_{k=0}^{N-1}\alpha_k \delta(s-s_k)$ with
$s_0=0$ and $\alpha_0=0$ for definiteness. From Eq.(4) with $s=L$,
the expression of $q$ can be calculated in closed form and reads
\begin{equation}
q(\lambda)=\frac{L}{N}\exp \left(-i \pi \lambda_0 / \lambda \right)
\frac{\sin( \pi \lambda_0 / \lambda)}{\pi \lambda_0/\lambda}F( \pi
\lambda/ \lambda_0)
\end{equation}
 where we have set $F(\xi)=\sum_{k=0}^{N-1} \exp(-2i \xi \rho_k)$
 and $\rho_k=k+(a n_s / \lambda_0)(\alpha_0+\alpha_1+...\alpha_k)$.
Appropriate choice of the values $\alpha_k$ may flatten the shape of
$\Gamma(\lambda)$ at around its zero $\lambda=\lambda_0$. A first
approach consists of considering the Taylor expansion of
$\Gamma(\lambda)$ at around $\lambda_0$ and imposing $\Gamma=
(\partial \Gamma /
\partial \lambda)=...=(\partial^M \Gamma / \partial \lambda^M)=0$
at
 $\lambda=\lambda_0$ up to a certain order $M \geq 1$, which  defines the spectral flatness order.
 This requires $F(\pi)=F'(\pi)=...=F^{(M-1)}(\pi)=0$,
 which yields
 \begin{equation}
 \sum_{k=0}^{N-1}  \left(k+\frac{a n_s}{\lambda_0}\sum_{j=0}^k \alpha_j \right)^l\exp \left(-2
 \pi i
\frac{a n_s}{\lambda_0} \sum_{j=0}^k \alpha_j \right)=0
 \end{equation}
 \begin{figure}[htb]
 \centerline{\includegraphics[width=8.2cm]{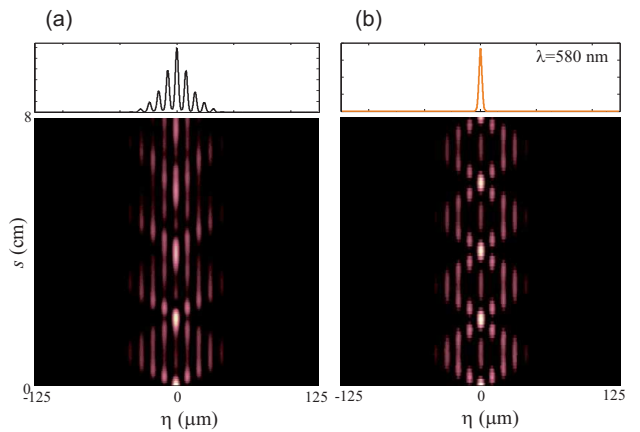}} \caption{
 (Color online) Intensity beam evolution (a) in the polychromatic, and (b) monochromatic (at carrier
 wavelength) regimes in a $L=8$-cm-long circularly-curved array without lumped phase slips.
 The insets at the top are the intensity profiles at the output
 plane. Parameter values are given in the text.}
 \end{figure}

\begin{figure}[htb]
\centerline{\includegraphics[width=8.2cm]{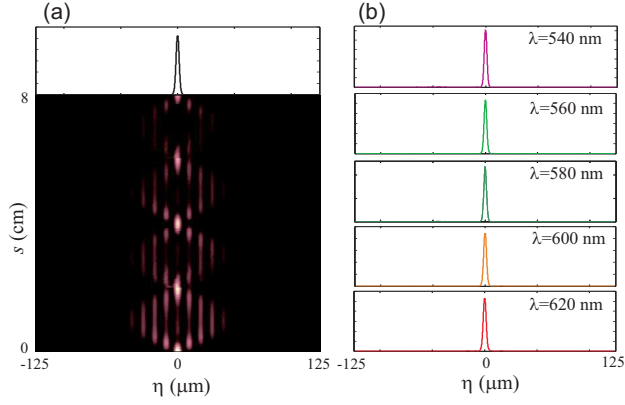}} \caption{
(Color online) Polychromatic BOs as obtained by insertion of two
lumped waveguide tilts at $s_1=2$ cm and $s_3=6$ cm (tilt angle
$\theta=\theta_B=25$ mrad). In (b) the output intensity profiles at
the output plane for the different spectral components are
depicted.}
\end{figure}

\begin{figure}[htb]
\centerline{\includegraphics[width=8.2cm]{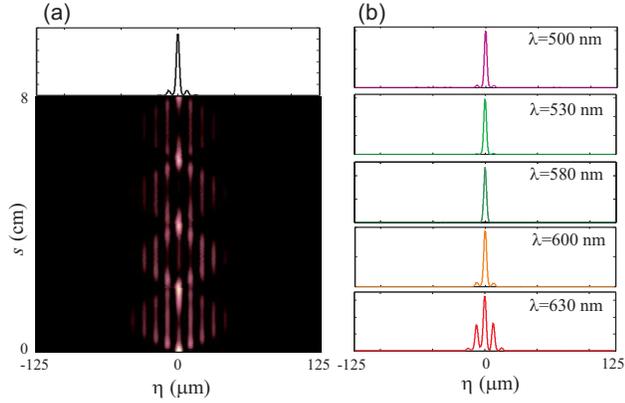}} \caption{
(Color online) Same as Fig.3, but with the insertion of three lumped
waveguide tilts at $s_1=2$ cm,  $s_2=4$ cm and $s_3=6$ cm (tilt
angle $\theta=2 \theta_B/3 \simeq $ 16.7 mrad).}
\end{figure}

for $l=0,1,..,M-1$. As an example, let us consider the case $N=4$,
previously discussed in Fig.2. Then one can check that Eq.(7) is
satisfied up to the order $M=2$ by simply assuming
$\alpha_1=\alpha_3=\lambda_0/(2 a n_s)$ and $\alpha_2=0$. This
corresponds to the introduction, at the arc lengths
$s_1=s_B(\lambda_0)$ and $s_3=3 s_B(\lambda_0)$, a lumped $\pi$
phase shift uniformly across the array. The corresponding behavior
of $\Gamma(\lambda)$ is shown by the dashed curve in Fig.1(c).
Polychromatic BOs and (approximate) self-imaging over the entire
spectral range (540,620) nm is demonstrated in Fig.3(a), where the
intensity evolution of the polychromatic beam of Fig.2(a) is now
computed with insertion of the phase gradients at
$s_1=s_B(\lambda_0)=2$ cm and at $s_3=3s_B(\lambda_0)=6$ cm. In the
numerical simulations, the lumped $\pi$ phase slips are simply
obtained by tilting the waveguides (tilting angle $\theta=\theta_B=
25$ mrad) as in \cite{Eisenberg00}. The spectrally-resolved output
intensity profiles of the polychromatic beam are also shown in
Fig.3(b). Note that, with this choice of $\alpha_1$, $\alpha_2$
 and $\alpha_3$, the function $\Gamma(\lambda)$ remains
 reasonably small in the range $\lambda \sim (540,620)$ nm, however
 it rapidly increases outside this interval. A different choice for $\alpha_1$, $\alpha_2$
 and $\alpha_3$, which broadens the spectral range where $2
 \Gamma(\lambda)$ remains smaller than $\sim 1$, is
 $\alpha_1=\alpha_2=\alpha_3=\lambda_0/(3 a n_s)$. In this case,
 $\Gamma'(\lambda_0)$ and $\Gamma^{''}(\lambda_0)$ do not vanish, however two additional zeros of $\Gamma$ are introduced at
 $\lambda \simeq 516$ nm and $\lambda \simeq 618$ nm, as shown by the dotted curve in Fig.1(c).
 The corresponding behavior of polychromatic BOs, for a beam
covering a spectral extent $\Delta \lambda \simeq 130$ nm, is shown
in Fig.4. Different algorithms, based on optimization methods (e.g.,
trying to minimize $\int d  \lambda |\Gamma(\lambda)|$), might be
investigated to flatten the function $\Gamma(\lambda)$ over a
prescribed spectral range. As a general rule,
 the increase of the number $N$ of phase gradients makes it the flattening procedure more flexible,
 and thus the polychromatic imaging of higher  quality  or applicable to a broader spectral range (for
 instance by increasing the spectral flattening order $M$).\\
In conclusion, broadband self-imaging based
 on polychromatic BOs in circularly-curved waveguide arrays has been
 proposed by insertion of lumped phase slips, which combat the dispersion of BO period for the
  various spectral beam components.

Author E-mail address: longhi@fisi.polimi.it

\end{document}